\documentstyle[prd,preprint,aps]{revtex}
\begin{document}
\draft
\title{Bounds on the magnetic moment and electric dipole moment\\
      of the $\tau$-neutrino via the process $e^{+}e^{-}\rightarrow \nu \bar \nu \gamma$}

\author{A. Guti\'errez-Rodr\'{\i}guez $^{1}$, M. A. Hern\'andez-Ru\'{\i}z $^{2}$
       and  A. Del Rio-De Santiago $^{1}$}
\address{(1) Facultad de F\'{\i}sica, Universidad Aut\'onoma de Zacatecas\\       
Apartado Postal C-580, 98068 Zacatecas, Zac. M\'exico.}

\address{(2) Facultad de Ciencias Qu\'{\i}micas, Universidad Aut\'onoma de Zacatecas\\
Apartado Postal 585, 98068 Zacatecas, Zac. M\'exico.}

\date{\today}
\maketitle
\begin{abstract}

Bounds on the anomalous magnetic moment and the electric dipole
moment of the $\tau$-neutrino are calculated through the reaction
$e^{+}e^{-}\rightarrow \nu \bar \nu \gamma$ at the $Z_1$-pole, and
in the framework of a left-right symmetric model. The results are
based on the recent data reported by the L3 Collaboration at CERN LEP.
We find that the bounds are almost independent of the mixing angle
$\phi$ of the model in the allowed experimental range for this parameter.
In addition, the analytical and numerical results for the cross-section
have never been reported in the literature before.
\end{abstract}

\narrowtext

\section{Introduction}
In many extensions of the Standard Model (SM) the neutrino acquires a nonzero
mass, a magnetic moment and an electric dipole moment \cite{Mohapatra}. In
this manner the neutrinos seem to be likely candidates for carrying features
of physics beyond the Standard Model \cite{S.L.Glashow}. Apart from masses
and mixings, magnetic moments and electric dipole moments are also signs of
new physics and are of relevance in terrestrial experiments, in the solar
neutrino problem, in astrophysics and in cosmology \cite{Cisneros}.

At the present time, all of the available experimental data for electroweak
processes can be understood in the context of the Standard Model of the
electroweak interactions (SM) \cite{S.L.Glashow}, with the exception of the
results of the SUPER-KAMIOKANDE experiment on the neutrino-oscillations \cite{fukuda},
as well as the GALLEX, SAGE, GNO, HOMESTAKE and LSND \cite{SAGE} experiments.
Nonetheless, the SM is still the starting  point for all the extended gauge
models. In other words, any gauge group with physical characteristics must have as a
subgroup the $SU(2)_{L} \times U(1)$ group of the standard model in such a way
that their predictions agree with those of the SM at low energies. The purpose
of the extended theories is to explain some fundamental aspects which are not
clarified in the frame of the SM. One of these aspects is the origin of the
parity violation at current energies. The Left-Right Symmetric Models (LRSM),
based on the $SU(2)_{R}\times SU(2)_{L}\times U(1)$ gauge group \cite{J.C.Pati},
give an answer to this problem by restoring the parity symmetry at high energies
and giving their violations at low energies as a result of the breaking of
gauge symmetry. Detailed discussions on LRSM can be found in the literature
\cite{R.N.Mohapatra,G.Senjanovic,G.Senjanovic1}.

In 1994, T. M. Gould and I. Z. Rothstein \cite{T.M.Gould} reported a bound on
the tau neutrino magnetic moment which they obtained through the analysis of
the process $e^{+}e^{-} \rightarrow \nu \bar \nu \gamma$, near the
$Z_1$-resonance, by considering a massive tau neutrino and using Standard     
Model $Z_1e^{+}e^{-}$ and $Z_1\nu \bar \nu$  couplings.

At low center of mass energy $s\ll M^{2}_{Z_1}$, the dominant contribution
to the process $e^{+}e^{-}\rightarrow \nu \bar \nu \gamma$ involves the
exchange of a virtual photon\cite{H.Grotch}. The dependence on the magnetic
moment comes from a direct coupling to the virtual photon, and the observed
photon is a result of initial state Bremsstrahlung. 

At higher $s$, near the $Z_1$ pole $s\approx M^{2}_{Z_1}$, the dominant 
contribution for $E_{\gamma} > 10$ $GeV$ \cite{L3A} involves the exchange of 
a $Z_1$ boson. The dependence on the magnetic moment and the electric dipole
moment now comes from the radiation of the photon observed by the neutrino
or antineutrino in the final state. The Feynman diagrams which give the most
important contribution to the cross section are shown in Fig. 1.
We emphasize here the importance of the final state radiation near the 
$Z_1$ pole, which occurs preferentially at high $E_{\gamma}$ compared to
conventional Bremsstrahlung.

Our aim in this paper is to analyze the reaction
$e^{+}e^{-}\rightarrow \nu \bar\nu \gamma$. We use recent data collected
with the L3 detector at CERN LEP \cite{L3A,L3B,ALEHP,L3} near the $Z_1$
boson resonance in the framework of a left-right symmetric model and we attribute
an anomalous magnetic moment and an electric dipole moment to a massive tau neutrino.
Processes measured near the resonance serve to set limits on the tau neutrino
magnetic moment and electric dipole moment. In this paper, we take
advantage of this fact to set bounds for $\mu_{\nu_{\tau}}$
and $d_{\nu_{\tau}}$ for different values of the mixing angle
$\phi$ \cite{M.Maya,J.Polak,L3C}, which is consistent with other constraints
previously reported \cite{T.M.Gould,H.Grotch,L3,A.Gutierrez,Aytekin} and
\cite{Escribano,Keiichi}.

We do our analysis near the resonance of the $Z_1$ ($s\approx M^{2}_{Z_1}$).
Thus, our results are independent of the mass of the additional heavy $Z_2$
gauge boson which appears in these kinds of models. Therefore,  we have the
mixing angle $\phi$ between the left and the right bosons as the only additional
parameter besides the SM parameters.

The L3 Collaboration evaluated the selection efficiency using detector-simulated
$e^{+}e^{-}\rightarrow \nu \bar \nu \gamma (\gamma)$ events, random trigger
events, and large-angle $e^{+}e^{-}\rightarrow e^+e^-$ events. A total of 14 events
were found by the selection. The distributions of the photon energy and the
cosine of its polar angle are consistent with SM predictions. The total
number of events expected from the SM is 14.1. If the photon energy is greater
than half the beam energy, 2 events are selected from the data and 2.4 events
are expected from the SM in the $\nu \bar\nu \gamma$ channel.

This paper is organized as follows: In Sect. II we describe the model with
the Higgs sector having two doublets and one bidoublet. In Sect. III we 
present the calculus of the process
$e^{+}e^{-}\rightarrow \nu \bar\nu \gamma$. In Sect. IV we make the
numerical computations. Finally, we summarize our results in Sect. V.

\section{The Left-Right Symmetric Model (LRSM)}

We consider a Left-Right Symmetric Model (LRSM) consisting of one bidoublet
$\Phi$ and two doublets $\chi _{L}$, $\chi _{R}$. The vacuum expectation
values of $\chi _{L}$, $\chi _{R}$ break the gauge symmetry to give mass to the left and right heavy
gauge bosons. This is the origin of the parity violation at low energies
\cite{R.N.Mohapatra}, {\it i.e.}, at energies produced in actual accelerators.
The Lagrangian for the Higgs sector of the LRSM is \cite{G.Senjanovic}

\begin{equation}
{\cal L}_{LRSM}=(D_{\mu}\chi _{L})^{\dagger}(D^{\mu}\chi _{L})
+ (D_{\mu}\chi _{R})^{\dagger}(D^{\mu}\chi _{R}) + Tr(D_{\mu}\Phi)^{\dagger}(D^{\mu}\Phi).
\end{equation}

\noindent The covariant derivatives are written as

\begin{eqnarray}
D_{\mu}\chi _{L}&=&\partial _{\mu}\chi _{L}-\frac{1}{2}ig{\bf \tau}\cdot {\bf W}_{L}\chi _{L}
-\frac{1}{2}ig^{'}B\chi _{L},\nonumber\\
D_{\mu}\chi _{R}&=&\partial _{\mu}\chi _{R}-\frac{1}{2}ig{\bf \tau}\cdot {\bf W}_{R}\chi _{R}
-\frac{1}{2}ig^{'}B\chi _{R},\\
D_{\mu}\Phi&=&\partial _{\mu}\Phi-\frac{1}{2}ig({\bf \tau}\cdot {\bf W}_{L}\Phi
-\Phi {\bf \tau}\cdot {\bf W}_{R}).\nonumber
\end{eqnarray}

In this model there are seven gauge bosons: the charged $W^{1}_{L,R}$, $W^{2}_{L,R}$
and the neutral $W^{3}_{L,R}$, $B$. The gauge couplings constants $g_L$ and $g_R$
of the $SU(2)_L$ and $SU(2)_R$ subgroups respectively, are equal: $g_{L} = g_{R}=g$,
since manifest left-right symmetry is assumed \cite{M.A.B}. $g'$ is the gauge
coupling for the $U(1)$ group.
 
The transformation properties of the Higgs bosons under the group $SU(2)_{L}\times SU(2)_{R}\times U(1)$
are $\chi _{L} \sim (1/2, 0, 1)$, $\chi _{R} \sim (0, 1/2, 1)$ and $\Phi \sim (1/2, 1/2^{*}, 0)$.
After spontaneous symmetry breaking, the ground states are of the form

\begin{equation}
\langle \chi _{L} \rangle = \frac{1}{\sqrt{2}}\left( 
					      \begin{array}{c}
					      0\\ 
					      v_{L}
					      \end{array}
					      \right), \hspace*{2mm}
\langle \chi_{R} \rangle = \frac{1}{\sqrt{2}}\left(                           
					     \begin{array}{c} 
					      0\\ 
					      v_{R}
					      \end{array}
					      \right), \hspace*{2mm}
\langle \Phi \rangle = \frac{1}{\sqrt{2}}\left(
					 \begin{array}{ll}
					 k&0\\
                                         0&k'
					 \end{array}
					 \right),
\end{equation}

\noindent breaking the symmetry group to form $U(1)_{em}$ giving mass
to the gauge bosons and fermions, with the photon remaining massless.
In Eq. (3), $v_L$, $v_R$, $k$ and $k'$ are the vacuum expectation values.
The part of the Lagrangian that contains the mass terms for the charged
boson is

\begin{equation}
{\cal L}^{C}_{mass} = (W^{+}_{L}\hspace*{2mm} W^{+}_{R}) M^{C}\left(
						\begin{array}{c}
						W^{-}_{L}\\
						W^{-}_{R}
						\end{array}
						\right),
\end{equation}

\noindent where $W^{\pm} = \frac{1}{\sqrt{2}}(W^{1} \mp W^{2})$.

The mass matrix $M^{C}$ is

\begin{equation}
M^{C} = \frac{g^{2}}{4}\left(
		       \begin{array}{cc}
		       v^{2}_{L}+k^{2}+k^{'2}&-2kk^{'}\\
		       -2kk^{'}&v^{2}_{R}+k^{2}+k^{'2}
		       \end{array}
                       \right).
\end{equation}

\noindent This matrix is diagonalized by an orthogonal transformation which is
parametrized \cite{M.A.B} by an angle $\zeta$. This angle has been restricted
to have a very small value because of the hyperon $\beta$ decay data \cite{M.Aquino}.

Similarly, the part of the Lagrangian that contains the mass terms for the neutral
bosons is

\begin{equation}
{\cal L}^{N}_{mass} = \frac{1}{8}(W^{3}_{L} \hspace*{2mm}  W^{3}_{R} \hspace*{2mm} B)
		      M^{N}\left(
			   \begin{array}{c}
			   W^{3}_{L}\\
			   W^{3}_{R}\\
			   B
			   \end{array}
			   \right),
\end{equation}

\noindent where the matrix $M^{N}$ is given by

\begin{equation}
M^{N} = \frac{1}{4}\left(
		   \begin{array}{ccc}
		   g^{2}(v^{2}_{L}+k^{2}+k^{'2})&-g^{2}(k^{2}+k^{'2})&-gg^{'}v^{2}_{L}\\
		   -g^{2}(k^{2}+k^{'2})&g^{2}(v^{2}_{R}+k^{2}+k^{'2})&-gg^{'}v^{2}_{R}\\
		   -gg^{'}v^{2}_{L}&-gg^{'}v^{2}_{R}&g^{'2}(v^{2}_{L}+v^{2}_{R})
		   \end{array}
		   \right).
\end{equation}

Since the process $e^{+}e^{-}\rightarrow \nu \bar \nu \gamma$ is neutral, we center our 
attention on the mass terms of the Lagrangian for the neutral sector, Eq. (6).

The matrix $M^{N}$ for the neutral gauge bosons is diagonalized by an orthogonal
transformation which can be written in terms of the angles $\theta _{W}$ and $\phi$ \cite{J.Polak1}

\begin{equation}
U^{N} = \left(
	\begin{array}{ccc}
	c_{W}c_{\phi}&-s_{W}t_{W}c_{\phi}-r_{W}s_{\phi}/c_{W}&t_{W}(s_{\phi}-r_{W}c_{\phi})\\
	c_{W}s_{\phi}&-s_{W}t_{W}s_{\phi}+r_{W}c_{\phi}/c_{W}&-t_{W}(c_{\phi}+r_{W}s_{\phi})\\
	s_{W}&s_{W}&r_{W}
	\end{array}
	\right),
\end{equation}

\noindent where $c_{W}=\cos \theta _{W}$, $s_{W}=\sin \theta _{W}$, $t_{W}=\tan\theta_{W}$
and $r_{W}=\sqrt{\cos 2\theta_{W}}$, with $\theta _{W}$ being the electroweak mixing angle.
Here, $c_{\phi}=\cos\phi$ and $s_{\phi}=\sin\phi$. The angle $\phi$ is considered
as the angle that mixes the left and right handed neutral gauge bosons $W^{3}_{L,R}$.
The expression that relates the left and right handed neutral gauge bosons $W^{3}_{L,R}$
and $B$ with the physical bosons $Z_{1}$, $Z_{2}$ and the photon is:

\begin{equation}
\left(
\begin{array}{c}
Z_{1}\\
Z_{2}\\
A
\end{array}
\right)
=
U^{N}\left(
     \begin{array}{c}
     W^{3}_{L}\\
     W^{3}_{R}\\
     B
     \end{array}
     \right).
\end{equation}

The diagonalization of (5) and (7) gives the mass of the charged
$W^{\pm}_{1,2}$ and neutral $Z_{1,2}$ physical fields:

\begin{equation}
M^{2}_{W_{1,2}}=\frac{g^{2}}{8}[v^{2}_{L}+v^{2}_{R}+2(k^{2}+k^{'2})
\mp \sqrt{(v^{2}_{R}-v^{2}_{L})^{2}+16(kk^{'})^{2}}],
\end{equation}

\begin{equation}
M^{2}_{Z_{1},Z_{2}}=B\mp \sqrt{B^{2}-4C},
\end{equation}

\noindent respectively, with

\[
B=\frac{1}{8}[(g^{2}+g^{'2})(v^{2}_{L}+v^{2}_{R})+2g^{2}(k^{2}+k^{'2})],
\]

\[
C=\frac{1}{64}g^{2}(g^{2}+2g^{'2})[v^{2}_{L}v^{2}_{R}+(k^{2}+k^{'2})(v^{2}_{L}+v^{2}_{R})].
\]

Taking into account  that $M^{2}_{W_{2}}\gg M^{2}_{W_{1}}$, from the expressions
for the masses of $M_{Z_{1}}$ and $M_{Z_{2}}$, we conclude that the relation
$M^{2}_{W_{1}}=M^{2}_{Z_{1}}\cos ^{2}\theta_{W}$ still holds in this model.

From the Lagrangian of the LRSM, we extract the terms for the neutral interaction
of a fermion with the gauge bosons $W^{3}_{L,R}$ and $B$:

\begin{equation}
{\cal L}^{N}_{int}=g(J^{3}_{L}W^{3}_{L}+J^{3}_{R}W^{3}_{R})+\frac{g^{'}}{2}J_{Y}B.
\end{equation}

Specifically, the interaction Lagrangian for $Z_{1}\rightarrow f\bar f$ \cite{P.Langacker} is

\begin{equation}
{\cal L}^{N}_{int}=\frac{g}{c_{W}}Z_{1}[(c_{\phi}-\frac{s^{2}_{W}}{r_{W}}s_{\phi})J_{L}    
		   -\frac{c^{2}_{W}}{r_{W}}s_{\phi}J_{R}],
\end{equation}

\noindent where the left (right) current for the fermions are

\[
J_{L,R}=J^{3}_{L,R}-\sin ^{2}\theta_{W}J_{em},
\]

\noindent and

\[
J_{em}=J^{3}_{L}+J^{3}_{R}+\frac{1}{2}J_{Y}
\]

\noindent is the electromagnetic current. From (13) we find that the amplitude
${\cal M}$ for the decay of the $Z_{1}$ boson with polarization $\epsilon ^{\lambda}$
into a fermion-antifermion pair is:

\begin{equation}
{\cal M}=\frac{g}{c_{W}}[\bar u \gamma^{\mu}\frac{1}{2}(ag_{V}-bg_{A}\gamma_{5})v]\epsilon^{\lambda}_{\mu},
\end{equation}

\noindent with

\begin{equation}
a=c_{\phi}-\frac{s_{\phi}}{r_{W}} \hspace{5mm} \mbox{and} \hspace{5mm}
b=c_{\phi}+r_{W}s_{\phi},
\end{equation}

\noindent where $\phi$ is the mixing parameter of the LRSM \cite{M.Maya,J.Polak}.

In the following section, we make the calculations for the reaction
$e^{+}e^{-}\rightarrow \nu \bar \nu \gamma$ by using the expression (14)
for the transition amplitude.

\section{The Total Cross Section}

We calculate the total cross section of the process $e^{+}e^{-}\rightarrow \nu \bar\nu \gamma$
using the Breit-Wigner resonance form \cite{Data2002,Renton}

\begin{equation}
\sigma(e^+e^-\rightarrow \nu \bar\nu \gamma)=\frac{4\pi(2J+1)\Gamma_{e^+e^-}\Gamma_{\nu \bar\nu \gamma}}{(s-M^2_{Z_{1}})^2+M^2_{Z_{1}}\Gamma^2_{Z_{1}}},
\end{equation}

\noindent where $\Gamma_{e^+e^-}$ is the decay rate of $Z_1$ to the channel
$Z_1 \to e^+e^-$ and $\Gamma_{\nu \bar\nu \gamma}$ is the decay rate of $Z_1$
to the channel $Z_1 \rightarrow \nu \bar\nu \gamma$.

In the next subsection we calculate the widths of Eq. (16).

\subsection{Width of $Z_1 \to e^+e^-$}

In this section we calculate the total width of the reaction

\begin{equation}
Z_1 \to e^+e^-,
\end{equation}

\noindent in the context of the left-right symmetric model which is described
in Section II.

The expression for the total width of the process $Z_1 \to e^+e^-$,
due only to the $Z_1$ boson exchange, according to the diagrams depicted in Fig. 1,
and using the expression for the amplitude given in Eq. (14), is:

\begin{equation}
\Gamma_{(Z_1 \to e^+e^-)}=\frac{G_FM^3_{Z_{1}}}{6\pi \sqrt{2}}\sqrt{1-4\eta}
[a^2(g^e_V)^2(1+2\eta)+b^2(g^e_A)^2(1-4\eta)],
\end{equation}

\noindent where $\eta=m^2_e/M^2_{Z_{1}}$.

\noindent We take $g^e_V=-\frac{1}{2}+2\sin^2\theta_W$ and $g^e_A=-\frac{1}{2}$
from the experimental data \cite{Data2002}, so that the total
width with $m_e=0$ is

\begin{equation}
\Gamma_{(Z_1 \to e^+e^-)}=\frac{\alpha M_{Z_{1}}}{24}
[\frac{\frac{1}{2}(a^2+b^2)-4a^2x_W+8a^2x^2_W}{ x_W(1-x_W)}],
\end{equation}

\noindent where $x_W=\sin^2\theta_W$ and $\alpha =e^2/4\pi$ is the fine structure
constant.

\subsection{Width of $Z_1 \rightarrow \nu \bar\nu \gamma$}

The expression for the Feynman amplitude ${\cal M}$ of the process $Z_1 \rightarrow \nu \bar\nu \gamma$
is due only to the $Z_1$ boson exchange, as shown in the diagrams in
Fig. 1. We use the expression for the amplitude given in Eq. (14) and assume
that a massive Dirac neutrino is characterized by the following phenomenological
parameters: a charge radius $\langle r^{2}\rangle$, a magnetic moment $\mu_{\nu_{\tau}}=\kappa \mu_B$
(expressed in units of the Bohr magneton $\mu_B$) and an electric dipole moment
$d_{\nu_\tau}$. Therefore, the expression for the Feynman amplitude ${\cal M}$
of the process $Z_1 \rightarrow \nu \bar\nu \gamma$ is given by

\begin{eqnarray}
{\cal M}_{a}&=&[\bar u(p_{\nu})\Gamma^{\alpha}\frac{i}{(\ell\llap{/}-m_{\nu})}
(-\frac{ig}{4\cos\theta_W}\gamma^{\beta}(a-b\gamma_{5}))v(p_{\bar\nu})]
\epsilon^\lambda_\alpha(\gamma) \epsilon^\lambda_\beta(Z_1)
\end{eqnarray}

\noindent and

\begin{eqnarray}
{\cal M}_{b}&=&[\bar u(p_{\nu})(-\frac{ig}{4\cos\theta_W}\gamma^{\beta}(a-b\gamma_{5}))
\frac{i}{(k\llap{/}-m_{\nu})}\Gamma^{\alpha}v(p_{\bar\nu})]
\epsilon^\lambda_\alpha(\gamma) \epsilon^\lambda_\beta(Z_1),
\end{eqnarray}

\noindent where

\begin{equation}
\Gamma^{\alpha}=eF_{1}(q^{2})\gamma^{\alpha}+\frac{ie}{2m_{\nu}}F_{2}(q^{2})\sigma^{\alpha \mu}q_{\mu}+ eF_{3}(q^{2})\gamma_5\sigma^{\alpha \mu}q_{\mu},           
\end{equation}

\noindent is the neutrino electromagnetic vertex, $e$ is the charge of the electron,
$q^\mu $ is the photon momentum and $F_{1,2,3}(q^2)$ are the electromagnetic
form factors of the neutrino, corresponding to charge radius, magnetic moment
and electric dipole moment, respectively, at $q^2=0$ \cite{Escribano,Vogel}.
While $\epsilon^{\lambda}_{\alpha}$ and  $\epsilon^{\lambda}_{\beta}$
are the polarization vectors of photon and of the boson $Z_1$, respectively.
$l$ ($k$) stands by the momentum of the virtual neutrino (antineutrino),
and the coupling constants $a$ and $b$ are given in the Eq. (15).

After applying some of the theorems of traces of the Dirac matrices and of
sum and average over the initial and final spins, the square of the matrix
elements becomes

\begin{equation}
\sum_s\mid{\cal M}_T\mid^2=\frac{g^2}{4\cos^2\theta_W}(\mu^2_{\nu_\tau}+d^2_{\nu_\tau})
[(a^2+b^2)(s-2\sqrt{s}E_\gamma)+a^2E^2_\gamma\sin^2\theta_\gamma].
\end{equation}

Our following step, now that we know the square of the Eq. (23) transition
amplitude, is to calculate the total width of $Z_1 \rightarrow \nu \bar\nu \gamma$:

\begin{equation}
\Gamma_{(Z_1 \rightarrow \nu \bar\nu \gamma)}=\int\frac{\alpha(\mu_{\nu_\tau}^2+d_{\nu_\tau}^2)}{96\pi^2M_{Z_{1}}x_W(1-x_W)}
[(a^2+b^2)(s-2\sqrt{s}E_\gamma)+a^2E^2_\gamma\sin^2\theta_\gamma]E_\gamma dE_\gamma d\cos\theta_\gamma,
\end{equation}

\noindent where $E_{\gamma}$ and $\cos\theta_{\gamma}$ are the energy and
scattering angle of the photon.

The substitution of (19) and (24) in (16) gives

\begin{eqnarray}
\sigma(e^{+}e^{-}\rightarrow \nu \bar\nu \gamma)&=&\int\frac{\alpha^2(\mu_{\nu_\tau}^2+d_{\nu_\tau}^2)}{192\pi}
[\frac{\frac{1}{2}(a^2+b^2)-4a^2x_W+8a^2x^2_W}{x^2_W(1-x_W)^2}]\nonumber\\
&&[\frac{(a^2+b^2)(s-2\sqrt{s}E_\gamma)+a^2E^2_\gamma\sin^2\theta_\gamma}{(s-M^2_{Z_{1}})^2+M^2_{Z_{1}}\Gamma^2_{Z_{1}}}]
E_\gamma dE_\gamma d\cos\theta_\gamma.
\end{eqnarray}

It is useful to consider the smallness of the mixing angle $\phi$,
as indicated in the Eq. (29), to approximate the cross section in
(25) by its expansion in $\phi$ up to the linear term:\\
$\sigma=(\mu_{\nu_\tau}^2+d_{\nu_\tau}^2)[A+B\phi+O(\phi^2)]$,
where $A$ and $B$ are constants which can be evaluated. Such an approximation
for deriving the bounds of $\mu_{\nu_\tau}$ and $d_{\nu_\tau}$ is more
illustrative and easier to manipulate.

For $\phi < 1$, the total cross section for the process $e^{+}e^{-}\rightarrow \nu \bar\nu \gamma$
is given by

\begin{equation}
\sigma(e^{+}e^{-}\rightarrow \nu \bar\nu \gamma)=(\mu_{\nu_\tau}^2+d_{\nu_\tau}^2)[A+B\phi+O(\phi^2)],
\end{equation}

\noindent where explicitly $A$ is:

\begin{eqnarray}
A=\int\frac{\alpha^2}{96\pi}
[\frac{1-4x_W+8x^2_W}{x^2_W(1-x_W)^2}]
[\frac{s-2\sqrt{s}E_\gamma+\frac{1}{2}E^2_\gamma\sin^2\theta_\gamma}{(s-M^2_{Z_{1}})^2+M^2_{Z_{1}}\Gamma^2_{Z_{1}}}]
E_\gamma dE_\gamma d\cos\theta_\gamma,
\end{eqnarray}

\noindent while $B$ is given by

\begin{eqnarray}
B&=&\int\frac{\alpha^2}{48\pi}
\{[\frac{3-8x_W}{r_Wx_W(1-x_W)^2}][\frac{s-2\sqrt{s}E_\gamma+\frac{1}{2}E^2_\gamma\sin^2\theta_\gamma}{(s-M^2_{Z_{1}})^2+M^2_{Z_{1}}\Gamma^2_{Z_{1}}}]\nonumber\\
&-&[\frac{1-4x_W+8x^2_W}{r_Wx_W(1-x_W)^2}]  [\frac{x_W(s-2\sqrt{s}E_\gamma)+\frac{1}{2}E^2_\gamma\sin^2\theta_\gamma}{(s-M^2_{Z_{1}})^2+M^2_{Z_{1}}\Gamma^2_{Z_{1}}}]\}
E_\gamma dE_\gamma d\cos\theta_\gamma.
\end{eqnarray}

The expression given for $A$ corresponds to the cross section previously
reported by T.M. Gould and I.Z. Rothstein \cite{T.M.Gould}, while $B$ comes
from the contribution of the LRSM.

Evaluating the limit when the mixing angle is $\phi=0$, the second term
in (26) is zero and Eq. (26) is reduced to the expression (3) given in
Ref. \cite{T.M.Gould}.

\section{Results}

In order to evaluate the integral of the total cross section as a function of
mixing angle $\phi$, we require cuts on the photon angle and energy to avoid
divergences when the integral is evaluated at the important intervals of each
experiment. We integrate over $\theta_\gamma$ from $44.5^o$ to $135.5^o$ and
$E_\gamma$ from 15 $GeV$ to 100 $GeV$ for various fixed values
of the mixing angle $\phi =-0.009, -0.005, 0, 0.004$. Using the following
numerical values: $\sin^2\theta_W=0.2314$, $M_{Z_1}=91.187$ $GeV$,
$\Gamma_{Z_1}=2.49$ $GeV$, we obtain the cross section
$\sigma=\sigma(\phi, \mu_{\nu_\tau},d_{\nu_\tau})$.

For the mixing angle $\phi$ between $Z_{1}$ and $Z_{2}$, we use the reported
data of M. Maya {\it et al.} \cite{M.Maya}:

\begin{equation}
-9\times 10^{-3}\leq \phi \leq 4\times 10^{-3},
\end{equation}

\noindent with a $90$ $\%$ C. L. Other limits on the mixing angle $\phi$ reported
in the literature are given in the Refs. \cite{J.Polak,L3C}.

As was discussed in Ref. \cite{T.M.Gould}, $N\approx\sigma(\phi, \mu_{\nu_\tau}, d_{\nu_\tau}){\cal L}$.
Using the Poisson statistic \cite{L3,Barnett}, we require that $N\approx\sigma(\phi, \mu_{\nu_\tau}, d_{\nu_\tau}){\cal L}$
be less than 14, with ${\cal L}= 137$ $pb^{-1}$, according to the data
reported by the L3 Collaboration Ref. \cite{L3} and references therein.
Taking this into consideration, we put a bound for the tau neutrino magnetic
moment as a function of the $\phi$ mixing parameter with $d_{\nu_\tau}=0$.
We show the value of this bound for values of the $\phi$ parameter in
Tables 1 and 2.\\

\begin{center}
\begin{tabular}{|c|c|c|}\hline
$\phi$&$\mu_{\nu_\tau} (10^{-6}\mu_B)$&$d_{\nu_\tau} (10^{-17}e \mbox{cm})$\\ \hline
\hline
-0.009&4.48&8.64\\
\hline
-0.005&4.44&8.56\\
\hline
0&4.40&8.49\\
\hline
0.004&4.37&8.43\\
\hline
\end{tabular}
\end{center}

\begin{center}
Table 1. Bounds on the $\mu_{\nu_\tau}$ magnetic moment and $d_{\nu_\tau}$
electric dipole moment for different values of the mixing angle $\phi$ before
the $Z_1$ resonance, {\it i.e.,} $s\approx M^2_{Z_1}$ .
\end{center}

These results are comparable with the bounds obtained 
in the references \cite{T.M.Gould,H.Grotch}.
However, the derived bounds in Table 1 could be improved by 
including data from the entire $Z_1$ resonance as is shown in Table 2.\\

\begin{center}
\begin{tabular}{|c|c|c|}\hline
$\phi$&$\mu_{\nu_\tau}(10^{-6}\mu_B)$&$d_{\nu_\tau}(10^{-17}e \mbox{cm})$\\ \hline
\hline
-0.009&3.37&6.50\\
\hline
-0.005&3.34&6.44\\
\hline
0&3.31&6.38\\
\hline
0.004&3.28&6.32\\
\hline
\end{tabular}
\end{center}

\begin{center}
Table 2. Bounds on the $\mu_{\nu_\tau}$ magnetic moment and $d_{\nu_\tau}$
electric dipole moment for different values of the mixing angle $\phi$ in
the $Z_1$ resonance, {\it i.e.,} $s=M^2_{Z_1}$ .
\end{center}

The above analysis and comments can readily be translated to the electric
dipole moment of the $\tau$-neutrino with $\mu_{\nu_\tau}=0$. The resulting
bound for the electric dipole moment as a function of the $\phi$ mixing
parameter is shown in Table 1.

The results in Table 2 for the electric dipole moment are in agreement
with those found by the L3 Collaboration \cite{L3}.

We end this section by plotting the total cross section in Fig. 2 as a function
of the mixing angle $\phi$ for the bounds of the magnetic moment given in
Tables 1, 2. We observe in Fig. 2 that for $\phi=0$, we reproduce the
data previously reported in the literature. Also, we observe that the
total cross section increases constantly and reaches its maximum value for
$\phi = 0.004$.

\section{Conclusions}

We have determined a bound on the magnetic moment and the electric dipole moment
of a massive tau neutrino in the framework of a left-right symmetric model as
a function of the mixing angle $\phi$, as shown in Table 1 and Table 2.

Other upper limits on the tau neutrino magnetic moment reported in the literature
are $\mu_{\nu_{\tau}} < 3.3 \times 10^{-6} \mu_{B}$  $(90 \% C.L.)$ from a sample
of $e^{+}e^{-}$ annihilation events collected with the L3 detector at the $Z_1$
resonance corresponding to an integrated luminosity of $137$ $pb^{-1}$ \cite{L3};
$\mu_{\nu_{\tau}} \leq 2.7 \times 10^{-6} \mu_{B}$ $(95 \% C.L.)$
at $q^2=M^2_{Z_1}$ from measurements of the $Z_1$ invisible width at LEP \cite{Escribano};
$\mu_{\nu_{\tau}} < 1.83 \times 10^{-6} \mu_{B}$ $(90 \% C.L.)$
from the analysis of $e^{+}e^{-}\rightarrow \nu \bar\nu \gamma$
at the $Z_1$-pole, in a class of $E_6$ inspired models with a light additional
neutral vector boson \cite{Aytekin}; from the order of
$\mu_{\nu_{\tau}} < O(1.1 \times 10^{-6} \mu_{B})$
Keiichi Akama {\it et al.} derive and apply model-independent bounds on the
anomalous magnetic moments and the electric dipole moments of leptons and quarks
due to new physics \cite{Keiichi}. However, the limits from Ref. \cite{Keiichi}
are for the tau neutrino with an upper bound of $m_\tau < 18.2$ $MeV$ which is
a direct experimental limit at present. It is pointed out in Ref. \cite{Keiichi}
however, that the upper limit on the mass of the electron neutrino and data from
various neutrino oscillation experiments together imply that none of the active
neutrino mass eigenstates is heavier than approximately 3 $eV$. In this case,
limits from Ref. \cite{Keiichi} improve by seven orders of magnitude. The limit
$\mu_{\nu_{\tau}} < 5.4 \times 10^{-7} \mu_{B}$ $(90 \% C.L.)$ is obtained
at $q^2=0$ from a beam-dump experiment with assumptions on the $D_s$
production cross section and its branching ratio into $\tau \nu_\tau$ \cite{A.M.Cooper},
thus severely restricting the cosmological annihilation scenario \cite{G.F.Giudice}.
Our results in Tables 1 and 2 confirm the bound obtained by the L3 Collaboration \cite{L3}.

In the case of the electric dipole moment, other upper limits reported in
the literature are \cite{Escribano,Keiichi}:

\begin{eqnarray}
\mid d(\nu_{\tau})\mid &\leq& 5.2 \times 10^{-17} \mbox{$e$ cm} \hspace*{4mm} {\mbox {95 \% C.L.},}\\
\mid d(\nu_{\tau})\mid &<& O(2 \times 10^{-17} \mbox{$e$ cm}). 
\end{eqnarray}

In summary, we conclude that the estimated bound for the tau neutrino magnetic
moment and the electric dipole moment are almost independent of the experimental
allowed values of the $\phi$ parameter of the model. In the limit $\phi = 0$,
our bound takes the value previously reported in the Ref. \cite{L3}.
In addition, the analytical and numerical results for the cross-section
have never been reported in the literature before and could be of relevance
for the scientific community.


\begin{center}
{\bf Acknowledgments}
\end{center}

This work was supported in part by {\bf SEP-CONACYT} (M\'exico)
({\bf Projects: 2003-01-32-001-057, 40729.F}), {\it Sistema Nacional de Investigadores}
({\bf SNI}) (M\'exico) and {\it Programa de Mejoramiento al Profesorado} ({\bf PROMEP}).
The authors would also like to thank Maureen Sophia Harkins Kenning for revising
the manuscript.

\newpage

\begin{center}
FIGURE CAPTIONS
\end{center}

\bigskip

\noindent {\bf Fig. 1} The Feynman diagrams contributing to the process $e^{+}e^{-}\rightarrow \nu \bar\nu \gamma$,
in a left-right symmetric model.

\bigskip

\noindent {\bf Fig. 2} The total cross section for $e^{+}e^{-}\rightarrow \nu \bar\nu \gamma$ 
as a function of $\phi$ and $\mu_{\nu_{\tau}}$ (Tables 1, 2).

\end{document}